\documentclass[a4paper]{jpconf}
\usepackage{graphicx}
\begin{document}
\begin{flushright}
SHEP-11-07
\end{flushright}

\title
{CP-violating Higgs sector of the MSSM 
through $gg\to H_1\to \gamma\gamma$}  

\author{
S. Hesselbach$^{a1}$
S. Moretti$^{b2}$
S. Munir$^{c3}$
P. Poulose$^{d4}$}

\address{ $^a$ GSI Helmholtzzentrum f\"ur Schwerionenforschung GmbH, Planckstrasse 1, 64291 Darmstadt, Deutschland}
\address{ $^b$ School of Physics \& Astronomy, University of Southampton,  Highfield, Southampton SO17 1BJ, UK}
\address{ $^c$ Department of Physics, COMSATS Institute of Information Technology, Defence Road, Lahore-54000, Pakistan}
\address{ $^d$ Physics Department, IIT Guwahati, Assam 781039, INDIA}

\ead{$^1$s.hesselbach@gsi.de}
\ead{$^2$stefano@soton.ac.uk}
\ead{$^3$smunir@ciitlahore.edu.pk}
\ead{$^4$poulose@iitg.ernet.in}  

\begin{abstract}
We study the effect of explicit CP violation in the Higgs sectors  of the MSSM
in the di-photon decay of the lightest CP-mixed Higgs state. Further
it is shown that the gluon fusion production mechanism along with the
di-photon decay enhances CP-violating effects for a large set of suitably chosen
parameter values.
\end{abstract}   

\section{Introduction}  
Phenomenology of the Minimal Supersymmetric Standard Model (MSSM) 
with CP-violating couplings can be very different from 
the one of CP-conserving MSSM. While spontaneous CP violation in the
MSSM arising from complex vacuum expectation values of the Higgs fields 
is essentially ruled out, many parameters in the MSSM, which
are absent in the Standard Model, can be complex, leading to large 
possibilities for CP violation. To avoid conflict with the low 
energy experimental constraints like those coming from measurements of the
Electric Dipole Moments (EDMs) of the electron, neutron and muon, the couplings 
relevant to the first and second generations are considered to be real.

By building on the results of  Refs.~\cite{Dedes:1999sj,Dedes:1999zh} (for the production) and \cite{Moretti:2007th}--\cite{Hesselbach:2007gf} (for the decay) -- see also Refs.~\cite{Dedes:2001zf}--\cite{Ellis:2004fs} -- we will look here at the Large Hadron Collider (LHC) phenomenology of the  $gg\to H_1\rightarrow \gamma\gamma$ process (where $H_1$ labels the lightest  neutral Higgs state of the CP-violating MSSM), which involves the (leading)  direct effects of CP violation through couplings of the $H_i$ ($i=1,2,3$  corresponding to the three neutral Higgs bosons)  to sparticles in the loops as well as the (sub-leading) indirect effects through  scalar-pseudoscalar mixing yielding  the CP-mixed state $H_i$.  (See Ref.~\cite{Hesselbach:2009st} for some preliminary accounts in this respect.) 
Here we summarize the results of \cite{Hesselbach:2009gw} focusing especially on the effects of a light stop in the production of a CP-mixed $H_1$ state by gluon fusion and its decay into two photons.

\section{CP violation in the di-photon Higgs search channel}  

Explicit CP violation arises in the Higgs sector of the MSSM when various related  couplings become complex. One consequence is that the physical Higgs bosons  are no more CP eigenstates, but a mixture of these \cite{Pilaftsis:1999qt}--\cite{Frank:2006yh}.  One may then look at the production and decay of the lightest of the physical Higgs  particles, hereafter labeled $H_1$. CP-violating effects in the combined production and decay process enter through complex $H_i$-$\tilde f$-$\tilde f^*$ couplings at the production and decay levels plus mixing in the propagator  ($\tilde f$ represents a sfermion).  

\begin{figure}[h!] 
\begin{center} 
\vskip 5cm
 \includegraphics{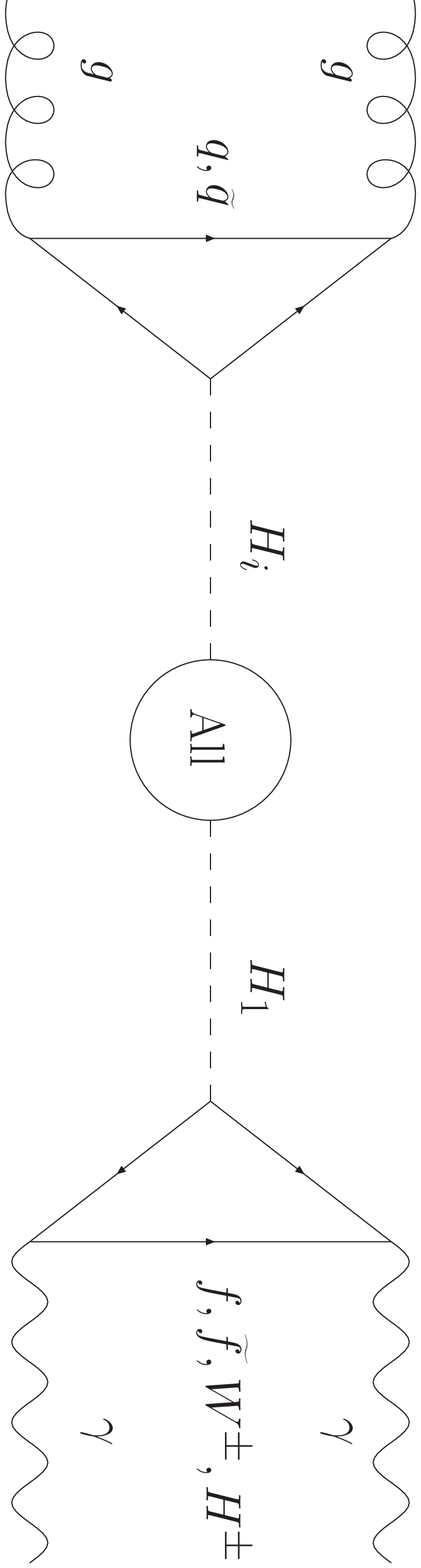}
\end{center} 
\caption{ Feynman diagram for $gg\rightarrow H_1\rightarrow \gamma\gamma$ including the  effect of mixing in the propagator. } 
\label{fig:feyn} \end{figure}

CP-conserving and  CP-violating effects enter at {\sl the same perturbative order} in the cross section for  $gg\to {\rm{Higgs}}\to \gamma\gamma$, so that the latter is an ideal laboratory to pursue studies of the complex (or otherwise) nature of the soft supersymmetry (SUSY) breaking parameters concerned. In contrast, notice that CP-violating effects through mixing in the propagator enter only at higher order, through self-energies, as there is already a tree-level CP-conserving contribution to the propagator\footnote{Also notice that there is also an `indirect' CP-violating contribution to the overall cross section of the process under study if one considers that the $H_1$ (and $H_{2,3}$) mass is subject to similar loop effects, though \cite{Moretti:2007th} has already shown that the consequent effects are marginal (see the left-hand side of Fig.~4 therein), so that  they are included here but not dwelt upon.}.   
The dynamics of CP-violating effects in the production and decay stages are rather similar, given that the same diagrammatic topologies are involved 
(see Fig.~\ref{fig:feyn}), 
in particular, as shown in our previous work \cite{Moretti:2007th}--\cite{Hesselbach:2007gf}, we expect a strong impact of a light stop quark in some regions of the parameter space.  
The propagator is considered in the following way. A Higgs particle, $H_i$, produced through gluon fusion,  can be converted into another mass eigenstate, $H_j$, through interaction of  fermion or gauge boson loops and their Supersymmetric counterparts (see Fig.~\ref{fig:feyn}).  Therefore,  in the following, when talking about results for the $H_1$, we consider the production of any of $H_i,~i=1,2,3$, which, while propagating, converts into $H_1$. 
However, whenever $M_{H_{2,3}}\le M_{H_1}+2$ GeV (assuming that mass measurements are resolved at about 2 GeV), we consider all the degenerate Higgs particles together.   

As intimated, the propagator matrix is obtained from the self-energy of the Higgs particles computed at one-loop level, where we used the expressions provided by \cite{Ellis:2004fs}, which include off-diagonal absorptive parts. The matrix inversion required is done numerically using the Lapack package \cite{lapack}. All relevant couplings and masses are obtained from CPSuperH  version 2 \cite{Lee:2007gn}, which takes into account all applicable experimental  constraints including the low energy EDMs. The cross section   of the full process shown in Fig.~\ref{fig:feyn} is computed numerically. The multi-dimensional integration is  carried out using the CUHRE program under the CUBA package \cite{Hahn:2004fe}. For our collider analysis we have used the CTEQ6 Parton Distribution Functions (PDFs) \cite{Pumplin:2002vw}--\cite{Lai:2007dq} computed at the factorization/renormalization scale $\mu=\sqrt{\hat{s}}$.  

\section{Results}  

In order to illustrate the typical effects of CP violation in the MSSM,  we have considered a few sample parameter space points and studied the effect of, in particular, light sparticles in the loops, chiefly, of stop squarks.  We fix the following MSSM parameters which play only a minor role in CP violation studies for Higgs production and decay:\\  

\(M_1=100~{\rm GeV},~~~M_2=M_3=1~{\rm TeV},~~~M_{Q_3}=M_{D_3}=M_{L_3}=M_{E_3}=M_{\rm SUSY}= 1~{\rm TeV}.\)\\  

We consider the case of all the third generation trilinear couplings being  unified into one single quantity, $A_f$. All the soft masses are taken to be the same at some unification scale, whose representative value adopted here is 1 TeV. When considering the light stop case we take a comparatively light value for $M_{U_3}\sim 250$ GeV, which corresponds to a stop mass of around 200 GeV,  otherwise $M_{U_3}$ is set to 1 TeV.  We could, alternatively, consider small values for $M_{Q_3}$ to reach light squarks, but the effects would qualitatively be the same. So, in   the following we keep a fixed value of $M_{Q_3}=1 $ TeV. In the Higgs scalar-pseudoscalar mixing the product of $\mu A_f$ is relevant rather than $\mu$ or $A_f$ separately. As argued in our earlier works, the only phase that is relevant is thus the sum of the phases of $\mu$ and $A_f$. In our analysis we have kept $\phi_{A_f}=0$ and studied the effect of CP violation by varying $\phi_\mu$. Regarding the absolute 
 values of $\mu$ or $A_f$, in our numerical analysis, we have varied these parameters between 1 and 1.5 TeV. $M_{H^+}$ is instead varied between 100 and 300 GeV. The mass of the lightest Higgs particle is consequently in the range of 50--130 GeV.  We then analyze cases with different values of $\tan\beta$. In particular, low $\tan\beta$ values give very small deviations from the corresponding  CP-conserving cases,  while large $\tan\beta$ values produce significant differences. Also, we take a  representative value of $\tan\beta=20$ to see the effect of the other parameters. 

\begin{figure}[h] \vskip 8cm \includegraphics{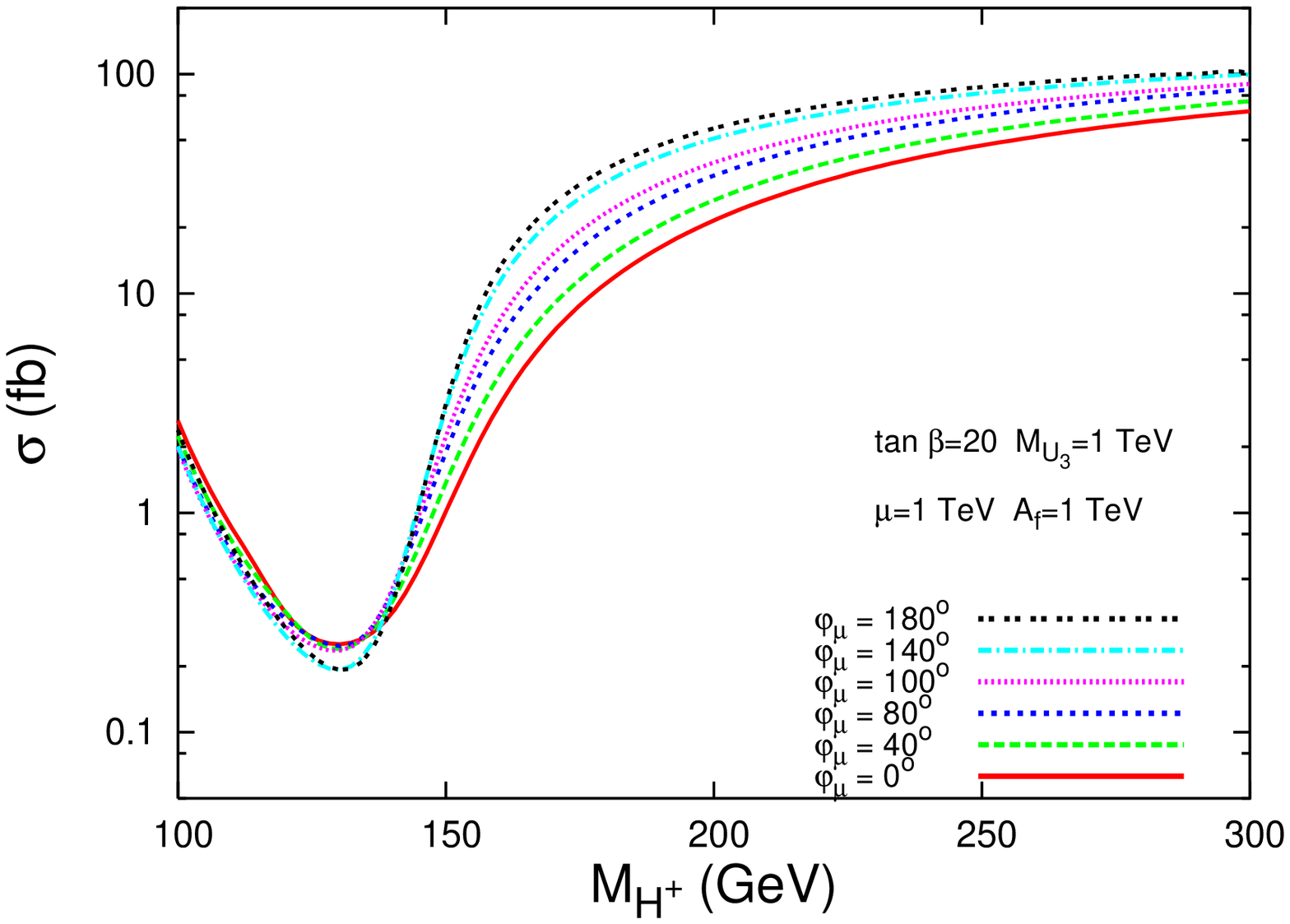} \includegraphics{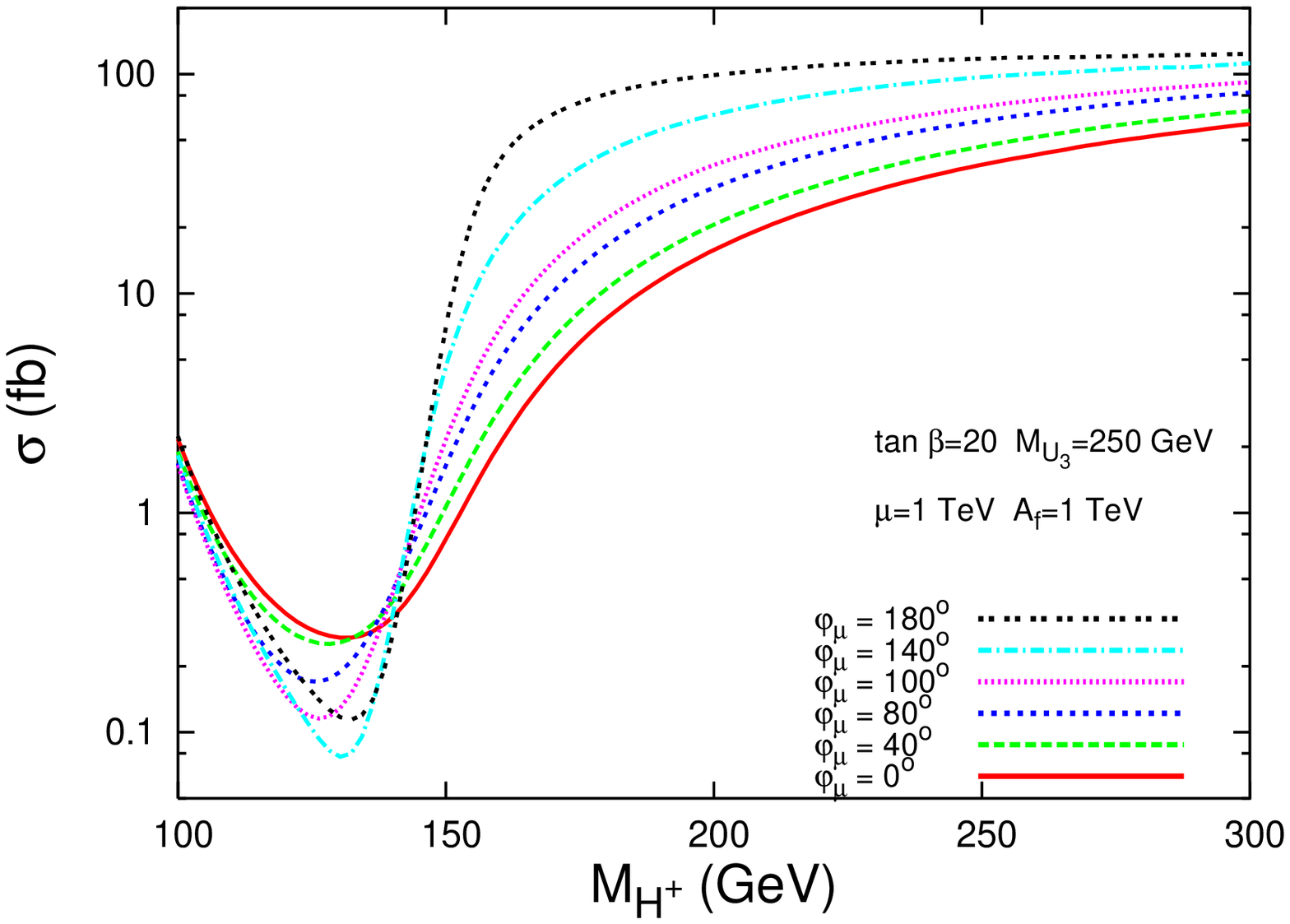} 
\caption{The cross section $\sigma (pp\rightarrow H_1 \rightarrow \gamma\gamma)$ at the LHC plotted against the charged Higgs boson mass for different $\phi_\mu$ values. The relevant MSSM variables are set as follows: $\tan\beta=20$, $A_f=1$ TeV, $\mu=1$ TeV, with $M_{U_3}=1$ TeV (left plot) and $M_{U_3}=250$ GeV (right plot).}  
\label{fig:tb20A1mu1phimu} \end{figure}

In Fig.~\ref{fig:tb20A1mu1phimu}  we plot the full cross section for $gg\to H_1\to \gamma\gamma$  against $M_{H^+}$.
We have considered $\mu=1$~TeV, $A_f=1$~TeV and $\tan\beta=20$.  There is appreciable variation of the cross section with  $\phi_\mu$. Comparing the two cases of light and heavy stops, it is clear that the effect of the Higgs-stop-stop coupling is significant. Indeed, this was also noticed when we studied the di-photon decay \cite{Hesselbach:2007jf}.   

In addition to the stop mass dependence of the $gg\to H_1\to\gamma\gamma$ cross section, we have also studied how the latter varies with the masses of the other (s)particles entering the loops. However, in line with the results of  Refs.~\cite{Moretti:2007th}--\cite{Hesselbach:2007gf} for the case of the $H_1\to\gamma\gamma$ decay, we have found that their impact is largely negligible here too, no matter the value of the CP-violating phases. 

\section{Conclusions} 

Significant effect of CP violation in the process $gg\to H_1\to \gamma\gamma$ is seen for the parameter values considered in this study.
Varying the mass of  the stop squark from about 1 TeV to around 250 GeV has also a strong quantitative impact on  the cross section, possibly changing its dependence on the phases also  qualitatively. 
The discovery channel for the mass range of the Higgs boson considered 
thus seem to have the potential to also disentangle its CP nature within the
context of the MSSM.

\ack{SM is financially supported in part by NExT institute. PP thanks IITG for travel support.} 

  \end{document}